\documentclass[oldversion]{aa}
\usepackage{latexsym}
\usepackage{graphicx}
\usepackage{amssymb}

\begin{document}
\title{Multiwavelength studies of the gas and dust disc of
IRAS~04158+2805
\thanks{Based on data collected at the
Canada-France-Hawaii Telescope. The CFHT corporation is funded by
the governments of Canada and France, and by the University of
Hawaii. Based also on data collected at ESO/VLT during observation
program 68-C.0171.}}

\author{A. M. Glauser \inst{1}
\and F. M\'enard \inst{2} \and C. Pinte \inst{4} \and G. Duch\^ene
\inst{2}\inst{3} \and M. G\"udel \inst{1} \and J.-L. Monin \inst{2}
\and D. L. Padgett\inst{5}} \offprints{A. M. Glauser,
\email{glauser@phys.ethz.ch}}
 \institute{Paul Scherrer Institut,
CH-5232 Villigen PSI, Switzerland \and Laboratoire d'Astrophysique
de Grenoble, CNRS/UJF UMR 5571, BP 53, F-38041 Grenoble cedex 9,
France \and Astronomy Department, UC Berkeley, Berkeley CA
947209-3411 USA \and School of Physics, University of Exeter,
Stocker Road, Exeter EX4 4QL, United Kingdom \and Caltech-JPL/IPAC,
Mail Code 100-22, Pasadena, CA 91125}

\date{Received December 2007 / Accepted April 2008}

\abstract{We present a study of the circumstellar environment of
\object{IRAS~04158+2805} based on multi-wavelength observations and
models. Images in the optical and near-infrared, a polarisation map
in the optical, and mid-infrared spectra were obtained with
VLT-FORS1, CFHT-IR, and {\sc Spitzer}-IRS. Additionally we used an
X-ray spectrum observed with Chandra. We interpret the observations
in terms of a central star surrounded by an axisymmetric
circumstellar disc, but without an envelope, to test the validity of
this simple geometry. We estimate the structural properties of the
disc and its gas and dust content. We modelled the dust disc with a
3D continuum radiative transfer code, MCFOST, based on a Monte-Carlo
method that provides synthetic scattered light images and
polarisation maps, as well as spectral energy distributions. We find
that the disc images and spectral energy distribution narrowly
constrain many of the disc model parameters, such as a total dust
mass of $1.0-1.75\cdot 10^{-4}\,M_\odot$ and an inclination of
$62\degr-63\degr$. The maximum grain size required to fit all
available data is of the order of $1.6-2.8\,\mu m$ although the
upper end of this range is loosely constrained. The observed optical
polarisation map is reproduced well by the same disc model,
suggesting that the geometry we find is adequate and the optical
properties are representative of the visible dust content. We
compare the inferred dust column density to the gas column density
derived from the X-ray spectrum and find a gas-to-dust ratio along
the line of sight that is consistent with the ISM value. To our
knowledge, this measurement is the first to directly compare dust
and gas column densities in a protoplanetary disc.}


\maketitle

\section{Introduction}

Accretion discs are key elements in star and planet formation
scenarios. They provide the material for accretion leading to star
and planet building, they provide the energy and material for
launching jets, and they are the medium through which angular
momentum is transported away. Knowing their geometrical and physical
properties is important for understanding these processes and their
evolution.

Large-scale surveys have been performed to search for discs around
young low-mass pre-main sequence stars (e.g., Stapelfeldt et al.
2003, Schneider et al. 2005), the so-called T Tauri stars. A few
tens of discs have been imaged and, for many of them, images are
available over a progressively broader wavelength range, enabling
deeper studies of the disc properties.

However, the properties of discs around more massive stars and, of
concern here, around the lower mass late M dwarfs and brown dwarfs
remain poorly known, because images for these discs are still
extremely rare. As a consequence, our knowledge of the circumstellar
environment of these objects is based solely on spectral energy
distribution (SED) fitting.

In this paper we present a study of IRAS~04158+2805, a low-mass star
located near the substellar boundary. The classification of
IRAS~04158+2805 varies in the literature, with most authors agreeing
on a late spectral type, implying a very low stellar mass. The
recent paper by Luhman (2006) concludes that its spectral type is
consistent with a low mass star close to the brown dwarf limit.
IRAS~04158+2805 is located at a distance of 140pc, in the L1495~east
dark cloud, which is part of the Taurus molecular cloud complex. It
is surrounded by an extended reflection nebulosity seen in scattered
light. It propels a well-collimated, ionised atomic jet seen in
H$_{\alpha}$ that extends at least out to 60 arcsec to the north.
The object is probably located in the foreground of the large
reflection nebulosity illuminated by V892 Tau, one of the rare
Herbig Ae stars of the Taurus cloud, because the disc of
IRAS~04158+2805 appears in silhouette over the reflection
nebulosity. This is the only such case in Taurus to our knowledge.

In this paper, we apply an axisymmetric, inclined-disc model to our
data, fitting near-infrared images, near- and mid-infrared
photometry, mid-infrared spectroscopy, millimeter photometry, and
near-infrared polarimetry to derive disc geometry, disc mass, and
dust composition. The large size of the disc derived from the
near-infrared images (radius of approximately 8 arcseconds in the I
band, corresponding to 1120 AU at a distance of 140 pc) is
intriguing, especially around a star of such a low mass. We will
show that the data are compatible with an inclined disc without any
need to invoke a larger envelope. The paper is organised as follows.
Section \ref{observation} presents the observations which are
further discussed in section \ref{results}. In section \ref{model},
we discuss the modelling of the dust disc. We estimate the
structural model parameters of the circumstellar disc and we discuss
the quality and the uniqueness of the solution found in section
\ref{discussion}. We present our conclusions in section
\ref{conclusion}.

\section{Observations and data reduction}\label{observation}

\subsection{Optical imaging and polarimetry}
\subsubsection{Observations} \label{sec:obs}

IRAS~04158+2805 was observed on December 12, 2001 in the I-band with
the {\sc FORS1/IPOL} instrument. The weather conditions were good
and the seeing was measured between 0\farcs9 and $1''$ over the
observation period. The total field of view (FOV) of {\sc
FORS1/IPOL} is $6\farcm8\times 6\farcm8$ in the Standard Resolution
mode with a focal scale of $0\farcs2$/pixel. Polarimetry was
performed by inserting a Wollaston prism in the beam. The prism
splits the incident light beam into two separate beams of orthogonal
polarisation states, the so-called ordinary ($o$) and extraordinary
($e$) beams.  A stepped half-wave plate retarder was placed at the
entrance of the incident beam and was rotated by steps of 22.5
degrees. The separation of the two $o-$ and $e-$ beams on the CCD is
performed by the Wollaston prism and overlap of the two beams is
avoided by inserting a 9-slit focal mask. Each slit in the mask
provided a $\sim20''\times 6\farcm8$ FOV. Images of IRAS~04158+2805
were recorded at 8 retarder positions with an integration time of
3~minutes per frame. The images were then combined to yield the
Stokes parameters I, Q and U.

\subsubsection{Data reduction pipeline}

\begin{figure*}
\resizebox{\hsize}{!}{\includegraphics{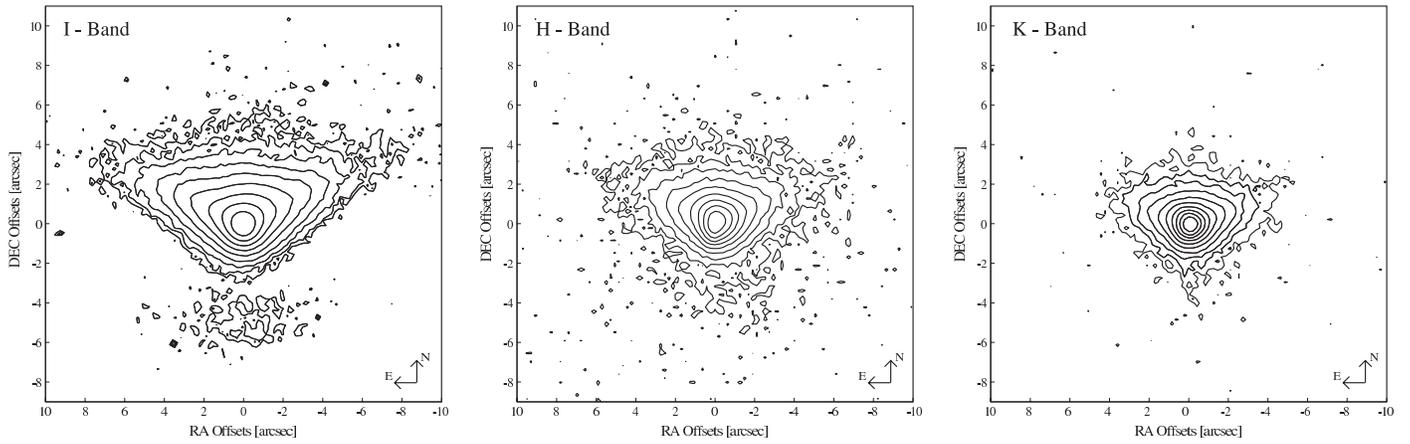}}
\caption{$I-$,$H-$ and $K-$band contour plots of IRAS~04158+2805.
The contour lines are on the levels $I_{\mathrm{max}}\cdot2^{-n}$
with $n=1,\dots,9$ or $10$ for the $K-$band image respectively.}
\label{contour}
\end{figure*}

A dedicated data reduction pipeline was written using {\sc
NOAO/IRAF} (see, e.g., Monin et al. 2006). The images are first
corrected for bias and bad pixels, and then flat-fielded. In the
next step, the images went through a polarisation extraction routine
in which the normalised flux difference between the ordinary and
extraordinary images is calculated for every pixel of the image, and
a Fourier series was computed to derive the Stokes parameter I, Q
and U\footnote{See the FORS user manual at http://www.eso.org. The
polarisation level, $P$, is obtained by calculating
$P=\sqrt{Q^2+U^2}/I$ and the position angle, $\Theta$, by
calculating $\Theta = 1/2\arctan(U/Q)$.}: We first computed the
normalised flux difference F from the target flux f:
\begin{equation}
F(\theta_i) = \frac{f^o(\theta_i) -f^e(\theta_i)}{f^o(\theta_i) +
f^e(\theta_i)}
\end{equation}
Thereafter we computed Q and U:
\begin{equation}
Q = \sum_{i=0}^{N-1} \frac{2}{N}F(\theta_i) \cos(4\theta_i) \\
U = \sum_{i=0}^{N-1} \frac{2}{N}F(\theta_i) \sin(4\theta_i)
\end{equation}
The absolute errors were estimated by using two independent methods:
first, from the photon noise on the $e-$ and $o-$beams separately,
and then propagating the errors in the calculations of Q, U, P and
$\Theta$; second, by measuring the standard deviation on the 8
images from the half-wave plate rotation. Both methods gave
consistent results, of order 0.3\%. The final intensity map is
presented in Fig.~\ref{contour}.

\subsubsection{Instrumental polarisation}

We estimated carefully the instrumental polarisation at the centre
of the FORS1 field by measuring nearby (i.e., high proper motion)
unpolarised targets. We observed GJ~781.1 and GJ~2147, two high
proper motion stars. Because the immediate solar neighbourhood is
remarkably devoid of dust (e.g., Leroy 1993, 1999) the interstellar
polarisation of nearby stars can be considered null. The average of
4 measurements on both GJ objects gives $P_{\rm inst}=0.02\%\pm
0.03\%$. We therefore believe that FORS1/IPOL instrumental
polarisation is very low on-axis, well below 0.1\% at the centre of
the field, and we did not attempt to remove it from the
measurements.

On the other hand, Patat \& Romaniello (2006) showed that FORS1
presents a spatially variable instrumental polarisation component.
This component follows a radial pattern with an intensity scaling as
$0.06\,r^2$ (in \% if $r$ is in arcmin), ranging from 0.1\% in the
central region, one arcmin in radius, to 1\% at the edge of the FOV.
First, the absolute value of the instrumental polarisation is low
(0.23\% at the position of the source) compared to the observed
polarisation of IRAS~04158+2805 which is always higher than 3\%.
Second, since the source is small compared to the FOV, we decided to
neglect the spatial variation of the instrumental polarisation on
the FORS1 detector because it does not vary significantly across the
object.

\subsection{Near-infrared imaging}

On October 29 and 30, 2001, we used the near-infrared CFHT-IR camera
(Starr et al. 2000) at the Canada-France-Hawaii Telescope to obtain
$H-$ and $K-$band images of IRAS~04158+2805 with a pixel scale of
0\farcs211/pixel and a total FOV of 3\farcm6. Conditions were
non-photometric and the seeing during the observations was 0\farcs65
at $K-$band and 0\farcs9 at $H-$band, as measured from the average
FHWM of several unresolved point sources located in the FOV. With
each filter, two series of 10 jittered images were obtained in
separate sets. Each set of images was first reduced as an
independent dataset in the following manner. All images were
median-combined to create a sky frame, which was subtracted from
each image. The images were then flat-fielded, registered based on
the location of a bright point source in the field and
median-combined. The two independent images per filter resulting
from this procedure were then averaged together to produce the final
images presented in Fig.~\ref{contour}.

\subsection{Mid-infrared spectroscopy}
We use archival data of the {\sc Spitzer} Infrared Spectrograph
(IRS, see, e.g., Houck et al. 2004) observation from March 4, 2004
(program request 3534848) which was done in the spectral mapping
mode with the two low resolution channels: Short-Low (SL; 5.2-14
$\mu$m, $\lambda/\Delta\lambda\sim90$) and Long-Low (LL; 14-38
$\mu$m, $\lambda/\Delta\lambda\sim90$). The mapping mode (3
exposures across the target for each slit and 2 nodding positions
each) was chosen due to the small mispointing of {\sc Spitzer} in
the early mission. Therefore, the flux of the object is separated in
two or three observations. To recover the photometric information,
all three parallel exposures were summed for the SL while the
overlap of the FOV of the LL-slit is such that the full flux is
extractable by summing the first and last exposures.

We used the final products from the Spitzer Science Center's IRS
data-reduction pipeline (post-BCD). To allow a background
subtraction the two nodding observations were used to subtract them
from each other. The data extraction was done with the Spitzer IRS
Customer Extractor software (SPICE) provided by the Spitzer Science
Center.

\subsection{X-ray spectroscopy}
\begin{figure}
\resizebox{\hsize}{!}{\includegraphics{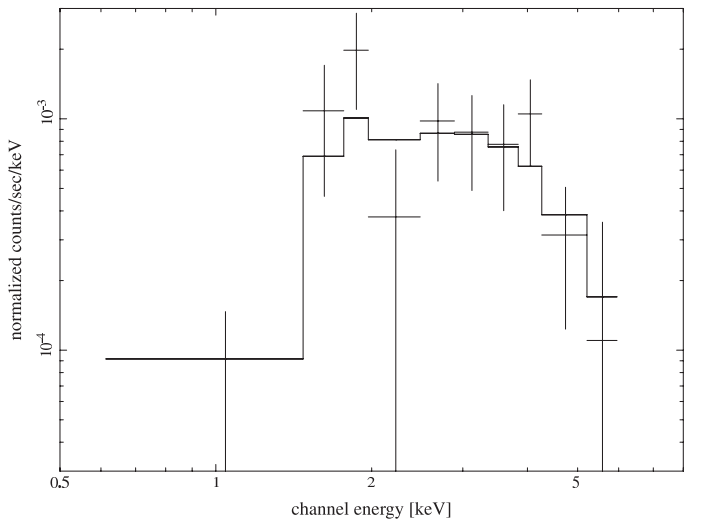}} \caption{X-Ray
spectrum of IRAS04158+2805 observed with Chandra} \label{xray}
\end{figure}

IRAS~04158+2805 was serendipitously observed by the Chandra X-ray
Observatory in a field pointing at the L1495 East dark cloud around
V410 Tauri. The observation was performed with ACIS-S on March 7,
2002 between 6:16~UT and 11:45~UT, with a useful total exposure time
of approximately 17700~s. IRAS 04158+2805 was located 11\farcm6
off-axis on the ACIS-S chip S1. This resulted in a rather distorted,
extended point spread function (PSF) with considerable background
contributions. The data were, however, sufficiently clean to extract
a useful spectrum (see Fig.~\ref{xray}).

We reduced all ACIS-S data using standard analysis techniques
following the Chandra CIAO Science
Threads\footnote{http://cxc.harvard.edu/ciao/threads/}.
Specifically, we flagged bad pixels, applied CCD charge transfer
inefficiency corrections to create a so-called level2 events file.
We extracted source counts from a circular area with a radius of
27\farcs6 around the source. To define a background spectrum, a
source-free circular area with a radius of 98\farcs4 was extracted
from the same chip. The counts were binned into a source and a
background spectrum. Appropriate responses were created using the
{\sl mkrmf} task, and the ancillary file was obtained from {\sl
mkarf}.

\section{Results}\label{results}

\subsection{Imaging}

Contour plots of IRAS~04158+2805 at I-, H- and K-bands are presented
in Fig.~\ref{contour}. In the I-band, the object shows a bipolar
reflection nebula geometry. The dark lane, tracing the equatorial
plane, is seen in absorption over the background light. It separates
a prominent triangular nebulosity located to the North from a low
surface brightness elongated counter nebula to the south. The
counter nebula is undetected at H- and K-bands. At these
wavelengths, the triangular shape of the main nebula is still
visible, but its extension decreases with increasing wavelength.

The maximum width of the northern nebula is measured to be
15\farcs8, 14\farcs1 and 10\farcs8 at I-, H- and K-band,
respectively. The northern nebula has a triangular shape whose
opening angle is $\sim$130\degr\ at all wavelengths. Finally, the
distance between the peak of the two nebulae at I-band is
4.8$\pm$0.2 arcsec. In Sect.~\ref{fitting}, we define additional
morphological and photometric observables to compare models and
observations.

\subsection{Aperture photometry}\label{aperture_photometry}

\begin{figure} \resizebox{\hsize}{!}{\includegraphics{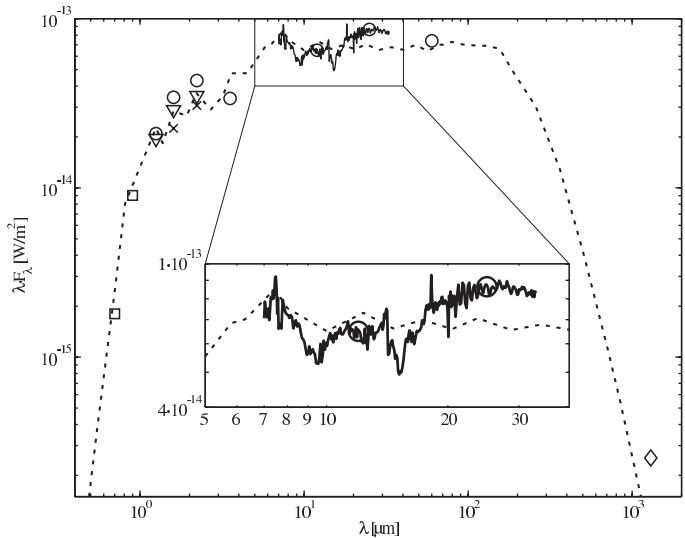}}
\caption{Modeled SED (dashed line) compared to observed fluxes: IRS
spectrum (Solid line); Strom~\&~Strom (1994, squares); Kenyon et al.
(1990) and Kenyon~\&~Hartmann (1995, circles); Luhman~\&~Rieke
(1998) and Luhman (2000, crosses); Motte~\&~Andr\'e (2001,
diamonds); this paper, see
sect.~\ref{aperture_photometry}~(triangle). The error bars of the
photometric fluxes are smaller than their symbols and therefore not
drawn.} \label{SED} \end{figure}

To construct the spectral energy distribution of IRAS~04158+2805
shown in Fig.~\ref{SED}, we use mid-infrared IRAS flux measurements
from Kenyon et al. (1990), optical and near-infrared photometry from
Kenyon et al. (1990), Strom \& Strom (1994) and from the 2MASS point
source catalog. The 1.3mm continuum flux is from Motte \& Andr\'e
(2001).
 Because IRAS~04158+2805 is
extended, we suspect that the 2MASS photometry may be underestimated
since these fluxes are listed in the 2MASS All-Sky Point Source
Catalogue (PSC) and were obtained by PSF fitting. To check this
point, we extracted the photometry from our own near-infrared images
by using three point-like sources that have 2MASS photometry and use
them as relative photometric standards. We obtained $K=11.04$ and
$H=12.10$ for IRAS~04158+2805, with estimated uncertainties of
0.05\,mag. For the H-band this is 0.28\,mag brighter than the 2MASS
photometry and 0.13\,mag for the K-band. While the brightness
profile of IRAS~04158+2805 is peaked in the near infrared and may
account for this discrepancy, we note that from our images we find a
difference of only 0.03\,mag in the photometry when two apertures of
4\farcs2 and 10\farcs6 are used. Therefore, we attribute the
difference in photometry of IRAS~04158+2804 between 2MASS and our
images to intrinsic variability of the source rather than an
aperture size too small in 2MASS.

\subsection{Mid-infrared spectroscopy}

The {\sc Spitzer}-IRS spectrum shows absorption features at
$\sim$10$\mu$m and $\sim$15.2$\mu$m. Both can be identified as
silicates and carbon dioxyde ice, respectively (see Fig.~\ref{SED}).
Similar features are seen for example in low-mass protostars, where
the cold envelope and/or outer disc cause the absorption (e.g.,
Watson et al. 2004). In IRAS~04158+2805, it is interesting to note
that the silicate absorption is maximum slightly short of 10
microns, around 9.5 microns. This is an indication for small dust
grains made of either amorphous or crystalline material such as
enstatite (Schegerer et al. 2006). We did not attempt to fit the
exact shape of the feature and cannot conclude about the exact
mineralogy of the grains. It is also interesting to find CO$_2$ ice
signatures in absorption in a source where the disc appears to
dominate the SED (see section \ref{sec_sed}), but the high
inclination, nearly edge-on, probably provides the needed column
density along the line of sight. However, it is beyond the scope of
this paper to discuss quantitatively the IRS spectrum. We will do so
in a forthcoming paper.

We extract the fluxes at 12$\mu$m and 25$\mu$m and compare with IRAS
fluxes published in Kenyon et al. (1990). The difference is of the
order of 0.5\% at 12$\mu$m and 1.3\% at 25$\mu$m and is negligible
therefore.

\subsection{I-Band polarimetric imaging}

\begin{figure*}\resizebox{\hsize}{!}{\includegraphics{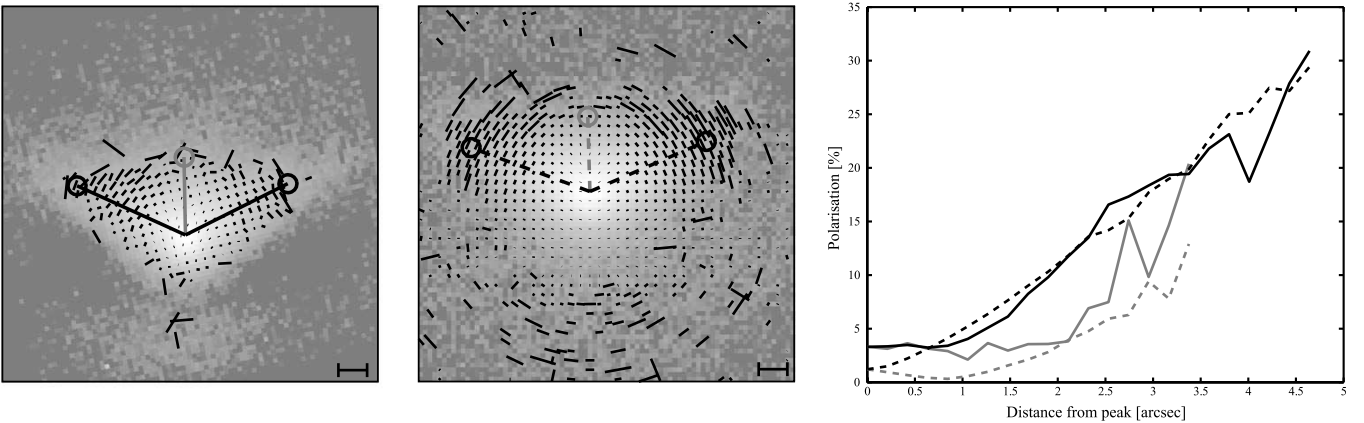}}
\caption{Polarisation maps of the observed ({\sl left panel}) and
modelled ({\sl middle panel}) nebula. The vector length for 100\%
polarisation is indicated by the bar in the lower right corner. {\sl
Right panel}: Comparison of the polarisation level as a function of
position in the observed (solid) and modelled (dashed) nebula. These
curves are estimated along the ridges (black curves) and symmetry
axis (gray) of the nebula.} \label{polarisation}
\end{figure*}

The light from the northern photometric nebulosity centre has a
linear polarisation of $3.3\pm1.2\%$, averaged within a region of 2
arcsec (10 pixels) in diameter. Since neighbouring objects do not
show a significant polarisation rate, this is an indication that the
peak in the nebula is not arising from stellar photons seen directly
but, instead, from scattered light. The polarisation is maximum at
the corners of the triangular northern nebula and has values between
35$\%$ and 41$\%$ (see Fig.~\ref{polarisation}). For this
determination we use the highest "reliable" value. We decided to use
only those pixels for which the polarisation rate topology shows a
certain smooth behaviour. This was quantified by determining the
standard deviation of the polarisation rate of each pixel and its 4
neighbours. If the standard deviation is higher than 20\% (typical
noise level), the pixel is not used.

\subsection{X-rays}\label{Result_Xray}

The Chandra ACIS X-ray spectrum of IRAS~04158+2805 is shown in
Fig.~\ref{xray}. The spectrum was binned such that the minimum
number of counts per bin was 10 before background subtraction. The
total number of counts in the 0.5-6 keV range was 100, which allowed
for only a rather simplistic spectral model that is, however,
sufficient to derive useful estimates of the absorption column
density to the X-ray source. The source spectrum is obviously very
strongly absorbed, with essentially no counts detected below
1.5~keV. The X-ray source is hard enough to produce counts up to 6
keV in the observation.

We fitted the spectral data in XSPEC (Arnaud et al. 1996), using the
{\sl vapec} collisional-ionization equilibrium model and a
photoelectric absorption component, the latter essentially being
defined by the hydrogen column density $N_H$ between observer and
source. We first fitted the spectrum with a single thermal
component, assuming that all element abundances are at 0.3 times the
solar values, referring to the solar abundances of Anders \&
Grevesse (1989). This corresponds to the elemental composition
typically measured in coronae of magnetically active stars (G\"udel
2004).

We found a best-fit temperature of 5.8~keV or $6.7\cdot 10^7$~K, and
a best-fit value for $N_H$ of $3.3^{+2.2}_{-1.4}\times
10^{22}$~cm$^{-2}$, where the error range refers to the 68\%
confidence limit. Given that the spectrum reveals only the hard end
of the entire soft X-ray spectrum, the fit may be biased by fitting
a hot component only. This bias may be relieved somewhat by fitting
a continuous distribution of emission measures (EMD\footnote{EMD
defines the the emission measure EM=$n_en_HV$ as a function of the
temperature, where $n_e,n_H$ are electron and hydrogen densities,
respectively, and V is the volume.}). We adopted an EMD with a
prescribed shape such as found for less-absorbed T Tauri stars,
described in more detail by Telleschi et al. (2007). The EMD model
essentially consists of two power laws on each side of a peak.
Neither the location of the EMD peak nor the power-law slope toward
higher temperatures were well constrained, but the results for $N_H$
were robust and converged to $N_H =
3.6^{+2.5}_{-1.0}\times10^{22}$~cm$^{-2}$.

To summarise these two approaches, irrespective of the uncertainties
in the intrinsic X-ray spectrum of IRAS~04158+2805, we find that the
X-rays, which are presumably emitted in the corona surrounding the
central star, are attenuated by a gas column density of $N_H \approx
3.5\times 10^{22}$~cm$^{-2}$, with a one sigma range of
$(1.9-6.1)\times 10^{22}$~cm$^{-2}$.

\section{Modelling of the dust disc}\label{model}

In this section we explore the parameter space of a simple model
comprising the stellar photosphere and a power-law disc (see
\ref{sys_geo} below) to explain the observations of IRAS~04158+2805
and its associated reflection nebulosity.

\subsection{Modelling technique}

Synthetic images, polarisation maps and spectral energy
distributions are computed using MCFOST, a 3D radiative transfer
code based on the Monte-Carlo method. MCFOST is described in details
in Pinte et al. (2006). MCFOST solves the full polarised radiative
transfer in dusty environment. It includes multiple scattering,
passive heating of the dust disc and thermal re-emission by the dust
to produce synthetic images in all four Stokes parameters at any
wavelength, as well as SEDs. The dust temperature, the same for all
grain sizes, is calculated assuming local thermal equilibrium.

\subsubsection{System geometry}\label{sys_geo}

We consider a flared density structure with a Gaussian vertical
profile $\rho(r,z) = \rho_0(r)\,\exp(-z^2/2\,h(r)^2)$, valid for a
well-mixed vertically isothermal, hydrostatic, non-self-gravitating
disc. We use power-law distributions for the surface density
$\Sigma(r) = \Sigma_0\,(r/r_0)^{\alpha}$ and for the scale height $
h(r) = h_0\, (r/r_0)^{\beta}$ where $r$ is the radial coordinate in
the equatorial plane and $h_0$ is the scale height at the radius
$r_0 =50$ AU. The disc extends from an inner cylindrical radius
$R_{\mathrm{in}}$ to an outer limit $R_{\mathrm{out}}$. The central
star is represented by a point-like, isotropic source of photons.

\subsubsection{Dust properties}

We consider homogeneous spherical grains and we use the dielectric
constants described by Mathis \& Whiffen (1989) in their model A,
which accounts for the interstellar extinction law. Grain sizes are
distributed according to a power-law $n(a) \propto a^{-3.7}$ with
$a_{\mathrm{min}}$ and $a_{\mathrm{max}}$ being the minimum and
maximum grain radii. The interstellar values from Mathis \& Whiffen
(1989) are $a_{\mathrm{min}}=0.005 \mu$m and $a_{\mathrm{max}}=0.9
\mu$m. The mean grain density is $0.5$ g~cm$^{-3}$ as a consequence
of the high porosity (80\%) of the grains. In this work,
$a_{\mathrm{max}}$ is considered as a free model parameter to fit
for. Extinction and scattering opacities, scattering phase functions
and Mueller matrices are calculated using Mie theory. The dust and
gas are assumed to be perfectly mixed and the grain properties are
taken to be independent of position within the disc.

\subsection{Data fitting}\label{fitting}

In the process of extracting the disc geometric parameters from the
observations, we first attempt to fit the I-, H- and K-band images
simultaneously with a single disc model. This is presented in
\ref{im_proc}. We use $a_{\mathrm{max}}$, $m_{\mathrm{dust\,disc}}$,
$\beta$, $\alpha$ and the inclination as free model parameters. The
I band image suggests a maximum disc radius of 8 arcsec, which
corresponds to 1120 AU at a distance of 140 pc. There must also be
an inner disc radius, which is typically at a few tenths of an AU.
These radii are poorly constrained by our I-, H, and K-band images;
we therefore fix them at the given values (inner radius is set to
0.5 AU), which will have little effect on our results. Since we do
not implement background illumination into our model, the appearance
of the dark lane and the counter nebula is expected to be slightly
different in the model.

We computed about 12000 models covering a wide range of the free
parameters. A handful of viable solutions remain after the
image-fitting process (in sec.~\ref{imagefitting}). We explore the
relative merits of these few solutions by calculating their
respective SEDs (in sec.~\ref{sec_sed}), and I-band polarisation map
(in sec.~\ref{sec_pol}) for comparison with the data.

\subsubsection{Image processing and extraction of observables}\label{im_proc}

To permit a direct comparison of the observed and modelled images,
observational effects must be added to the model images. First of
all, the pixel size of the model output is therefore chosen to match
the scale of the observation (0\farcs200/pixel in the optical,
0\farcs211/pixel in NIR). The model images must be convolved with
the observed PSF and scaled to the same peak value as the observed
images which allows an addition of Poisson and background noise to
yield the same signal-to-noise ratio.

\begin{figure}
\resizebox{\hsize}{!}{\includegraphics{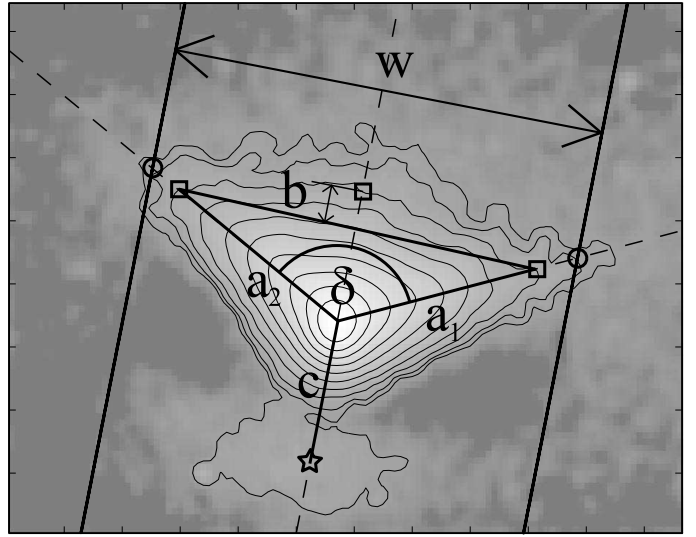}}
\caption{Definition of geometrical observables w,a,b,c and $\delta$.
The contour lines are shown for the levels $I_{\mathrm{max}}\cdot
2^{-n}$ $n=1\dots10$. The two ridges and the symmetry axis are drawn
as dashed lines. Their intersection points with the contour lines of
the flux level $I=I_{\mathrm{max}}\cdot2^{-8}$ are shown as boxes
and $I$=$1\cdot\sigma_\mathrm{noise}$ as circles, respectively. The
counter nebula peak is shown as a star.} \label{geometry}
\end{figure}

Having two comparable images, it is now possible to quantitatively
compare the model and observed images. To avoid potential biases
induced by asymmetries in the disc structure and the ignorance of
the background illumination of the object and to circumvent the
intrinsically noisy nature of Monte Carlo images, we decided to use
a ``quality metric'' based on geometrical observables of the nebula
rather than an image-wide pixel-to-pixel goodness-of-fit estimate.
These observables, which are detailed below and illustrated in
Fig.~\ref{geometry} are chosen to describe the width of the nebula,
its triangular shape and the counter nebula behaviour relative to
the northern nebula. They are extracted with the same routines in
the observed and modelled images, after further smoothing the images
with a gaussian (width $\sigma=1$~pixel).
We first define the outer ``ridges'' of the triangular nebula. To
allow this, we calculate contour lines at the flux levels
$I_n=I_{\mathrm{max}}\cdot2^{-n}$ with $n=1,2,\dots,N$, while N
defines the last contour line at flux level higher than the noise.
The ridges (labelled with $a_1$ and $a_2$ in Fig.~\ref{geometry})
are then defined by two lines that cross those points on each
contour that are most distant from the central peak. We derive these
lines as follows: We represent the contours in polar coordinates as
$r_n(\phi)$ where $r$ is the distance from the central peak and
$\phi$ is the angle from North. These polar functions are then
normalised and averaged ($r(\phi)=n^{-1}\sum_n
r_n(\phi)/\int_0^{2\pi} r_n(\phi')d\phi'$). The direction of the two
lines are described by the two values of $\phi$ where $r(\phi)$ has
a maximum. Thereafter, the opening angle $\delta$ between the two
ridges was measured. The symmetry axis of the nebula corresponds to
their bisecting line.

The width of the nebula is defined as the distance $w$ between the
two intersection points of the ridges with the contour at the level
of a signal-to-noise ratio equal to 1 (shown in Fig.~\ref{geometry}
as circles).

The next observable describes quantitatively whether the nebula
looks triangular or round. This observable, which we call the
triangularity, $t$, is determined for each contour level. In
Fig.~\ref{geometry} an example is shown for the contour level $n=8$.
The two ridges and the symmetry axis are intersected with the
contour line (shown as squares in the figure). The distance between
the intersection point of the symmetry axis with the contour line
and the triangle spanned by the peak and the two intersection points
of the contour line with the ridges is called $b$. The triangularity
is then defined as
\begin{equation}
t=\frac{b}{a(1-\cos(\delta/2))}
\end{equation}
where $a$ is the average of $a_1$ and $a_2$. This parametrisation
was chosen to describe a perfectly triangular nebula for $t=0$ and a
perfect circle for $t=1$. For the model-to-data comparison, we
fitted the most triangular contour line, which describes best the
shape of the outer nebula.

Finally, the counter nebula is described by the peak-to-peak
distance $c$, and the contrast is defined as the ratio between the
maximum brightness of the nebula and the counter nebula.

\subsubsection{Image fitting}\label{imagefitting} For a given set
of free model parameters ($\alpha$, $\beta$, ...), we use the
following quality metric to estimate whether the model actually
reproduces the three images of the object. This metric is a
pseudo-$\chi^2$, defined as follows: \begin{equation}\label{f2}
f^2=\sum_{\lambda,i}
\frac{(o^{model}_{i,\lambda}-o^{obs}_{i,\lambda})^2}{w_{i,\lambda}}\cdot
g_{i,\lambda} \end{equation} where $o^{model}_{i,\lambda}$ and
$o^{obs}_{i,\lambda}$ are the observables with index i (i.e. $w$,
$t$, $c$, $\delta$, contrast) for a given photometric band $\lambda$
(I, H, K) of the model image and observation image, respectively.
The normalisation with $w_{i,\lambda}$, which is determined by the
standard deviation
$std((o^{model}_{i,\lambda}-o^{obs}_{i,\lambda})^2)$ over the whole
explored model parameter space, allows an equitable comparison of
different quantities. Additionally, each observable is weighted with
$g_{i,\lambda}$. These weights were fine-tuned by hand to equalise
the influence of each individual parameter. This weighting is
important to allow for a wavelength-dependent contrast parameter
(e.g. the counter nebula is not visible in the H and K band,
therefore the parameters $c$ and the contrast are useless for these
bands). Additionally the use of some parameters provides stronger
constraints towards a good fitting solution than others (e.g. the
nebula width and triangularity in comparison to the opening angle).
The weights are listed in Table \ref{weights}. \begin{table}[h]
\caption{Weights $g_{i,\lambda}$ of the pseudo-$\chi^2$ function
(\ref{f2})} \label{weights} \centering \begin{tabular}{l|c c c c c}
\hline\hline &$\delta$&$w$&$t$&$c$&Contrast\\ \hline
I&1/9&3/9&3/9&1/9&1/9\\ H&1/7&3/7&3/7&0&0\\ K&1/7&3/7&3/7&0&0\\
\hline \end{tabular} \end{table}

\begin{figure}
\resizebox{\hsize}{!}{\includegraphics{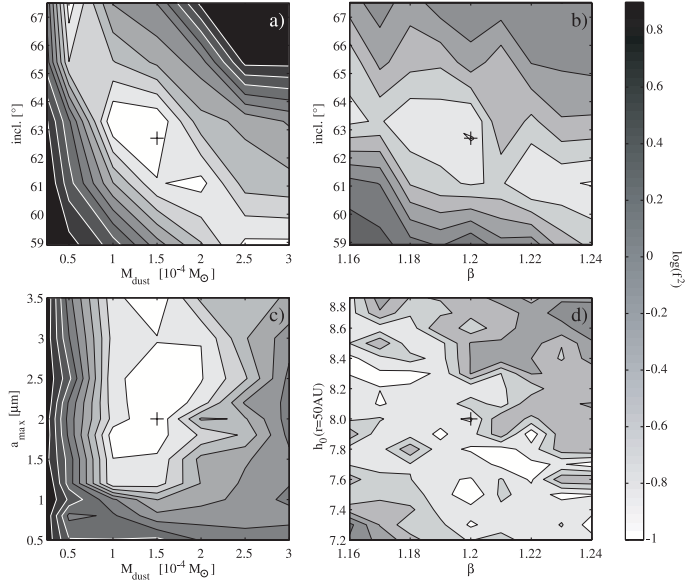}}
\caption{Examples of projections of $f^2$ (eqn. \ref{f2}) to
different model parameter planes. The contour is plotted in a
logarithmic scale. White areas correspond to minimal values of
$f^2$. The colour range is identical in all four plots. The black
cross shows the location of the best model, after the SED fitting.}
\label{chi2}
\end{figure}

The exploration of the parameter space was performed iteratively. We
first ran a series of models with only 2 variable model parameters.
Once the best model in the series was identified, we fixed these
model parameters and created a new series by varying two other
parameters, and so on. The range of the explored parameter space is
listed in the second column of Table~\ref{Bestfit}. The first
selection of the parameter range was chosen arbitrarily but
reasonably. If the model behaviour suggested possible solutions
outside the scanned range, the exploration was extended until the
model diverged significantly from the observation. Among all these
models, only a few provided good fits simultaneously to the images
of IRAS~04158+2805. In addition, we note that fitting the images at
each wavelength independently leads to models of similar parameters
and quality. The combination of all images in a single fit allows,
however, to narrow down the list of possible good models.

\begin{figure}
\resizebox{\hsize}{!}{\includegraphics{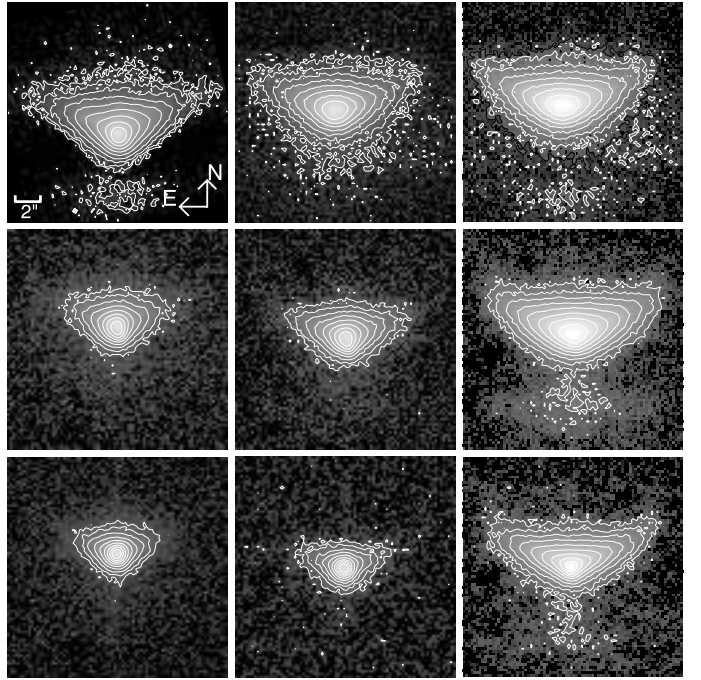}}
\caption{Comparison between observations (left) and best fitting
model (mid, see Table~\ref{Bestfit}) in I-band (top), H-band
(middle) and K-band (bottom). For demonstration purposes, the right
column shows the images of the best fitting model but with doubled
dust disc mass ($3\cdot 10^{-4}M_\odot$).} \label{Comp_im}
\end{figure}

Fig. \ref{chi2} shows the map of $f^2$ as a function of pairs of
free model parameters: Dust disc mass vs. inclination (\ref{chi2}a),
flaring coefficient $\beta$ vs. inclination (\ref{chi2}b), dust disc
mass vs. maximum grain size (\ref{chi2}c), $\beta$ vs. scale height
h$_0$ (\ref{chi2}d). These plots demonstrate the complexity of the
search for the minimum since some model parameters are strongly
correlated (shown, e.g., in Fig. \ref{chi2}a and b) or show numerous
local minima (e.g., Fig. \ref{chi2}d). While Fig. \ref{chi2}c
implies a smooth topology for the $M_{dust}$ vs. $a_{\mathrm{max}}$
plane, the minimum found at $M_{dust}=1.5\cdot10^{-4}M_\odot$ and
$a_{\mathrm{max}}=2\mu m$ is only valid for the inclination of
$i=63\degr$. At an inclination of $i=61\degr$ we find a minimum with
a similar goodness-of-fit with $M_{dust}=2.5\cdot10^{-4}M_\odot$ and
$a_{\mathrm{max}}=2.5\mu m$. This implies that we cannot find a
unique best model by fitting images only. The synthetic images for
our best-fitting model, found after the SED fitting process, is
presented in Fig.~\ref{Comp_im} alongside the observed images.

We estimated the valid model parameter range by using one
good-fitting parameter set and changed only one parameter until the
$f^2$ function showed a significant increase or until an eyeball
comparison could clearly detect an unsatisfying model. This method
does not include an estimation of the general solution range but
gives an impression of the validity of the free model parameters at
a certain minimum of $f^2$.

Additionally we had to set an artificial upper limit for the valid
range of $a_{\mathrm{max}}$: Fig. \ref{chi2}c shows a valley for
$M_{dust}=1.5\cdot10^{-4}M_\odot$ and $a_{\mathrm{max}}\ge2\mu m$.
This is a consequence of comparing the modelled with the observed
images only up to the K-band. Therefore, the fit is only mildly
sensitive to larger grains, and we set $a_{\mathrm{max}} = 2 \mu m$
with no further consequences for the solution.

Fitting the images for each wavelength individually delivers only
very small variations in $a_{\mathrm{max}}$ and disc mass and is
consistent with the multi-wavelength solution within the valid model
parameter range.

\subsubsection{SED fitting}\label{sec_sed}

To refine the model selection we fit the SED. The modelled SED
includes photospheric emission, scattered light and thermal emission
by assuming passive heating of the dust in the disc (Pinte et al.
2006). We use a 3000K spectrum from Baraffe et al. (1998) in
agreement with a spectral type of a low mass star. For each of the
best image-fitting models there are only 2 extra free parameters to
the SED fitting: the value of foreground extinction A$_V$ and the
bolometric luminosity of the central source. The latter is a free
parameter since we do not see the source directly. The foreground
extinction A$_V$ describes the material between the observer and the
outer limit of the star-disc system and is added to the model after
the radiation transfer calculation.

It turns out that only one model is a simultaneous good fit to the
images and to the SED. We consider it our best model in the
following (see third column of Table~\ref{Bestfit}). The SED for
this model is presented in figure \ref{SED}. It includes A$_V=0.5$
mag and a stellar luminosity of $\sim$0.4 $L_\odot$.

Photometric variations are observed for IRAS~04158+2805. It results
in a large spread in the SED data at a given wavelength in the
optical and NIR (e.g., $\sim$1.5mag in the I-band between Strom \&
Strom 1994 and Luhman 2000). The calculated SED agrees well with the
data points and falls within the range of observed variations. The
estimated extinction is lower than the value estimated by Strom \&
Strom (1994; A$_J$=0.7), Luhman \& Rieke (1998; A$_J$=1.5), and
Luhman (2000, A$_H$=1.5) in the NIR. However these estimations did
not take into account the fact that the object's colours are
strongly modified by the scattering in the circumstellar material.
We therefore believe that our estimate of the {\sl foreground}
extinction is more robust than previous estimates.

\subsubsection{Acceptable parameter range} The
parameters of the best-fitting model are described in
Table~\ref{Bestfit} under "best model". To define the "acceptable
range" (see fourth column of Table~\ref{Bestfit}) we calculated
series of images from the best model in which only one parameter at
a time was varied from its best fit value. This allowed us to define
a possible range of solutions, i.e., solutions with values of $f^2$
only moderately higher than the best model ($f^2/f_{min}^2\lesssim
125\%$) or images that showed no obvious mismatches with the
observations by an eyeball comparison. These ranges give a feeling
of how tightly each parameter is constrained.

\begin{table}[h] \caption{Disc model parameters for the
best-fitting model and their acceptable range.} \label{Bestfit}
\centering \begin{tabular}{l c c c c} \hline\hline Parameter &
Explored&Best&Acceptable\\ &range&model&range\\
 \hline
$a_{\mathrm{max}}\ [\mu m]$&0.5 - 4.0&2.0&1.6 - 2.8\\
$M_{\mathrm{dust\,disc}}\ [10^{-4} M_\odot]$ &0.25 - 3.0&1.5&1.0 - 1.75\\
$R_{\mathrm{out}} [AU]$ &not varied&1120&-\\
$R_{\mathrm{in}} [AU]$ &not varied&0.5&-\\
$\beta$ &1.0 - 1.3&1.20&1.15 - 1.21\\
$\alpha$ &-2.0 - -0.1& -1.0&-1.5 - -0.5\\
$h_0(r=50 AU) [AU]$&5.0 - 12.0&8.0&7.8 - 8.2\\
$incl. [\degr]$ &0.0 - 90.0&62.7&62.1 - 63.25\\
\hline
\end{tabular}
\end{table}

\subsubsection{Using polarisation to confirm the
model}\label{sec_pol}

\begin{figure}
\resizebox{\hsize}{!}{\includegraphics{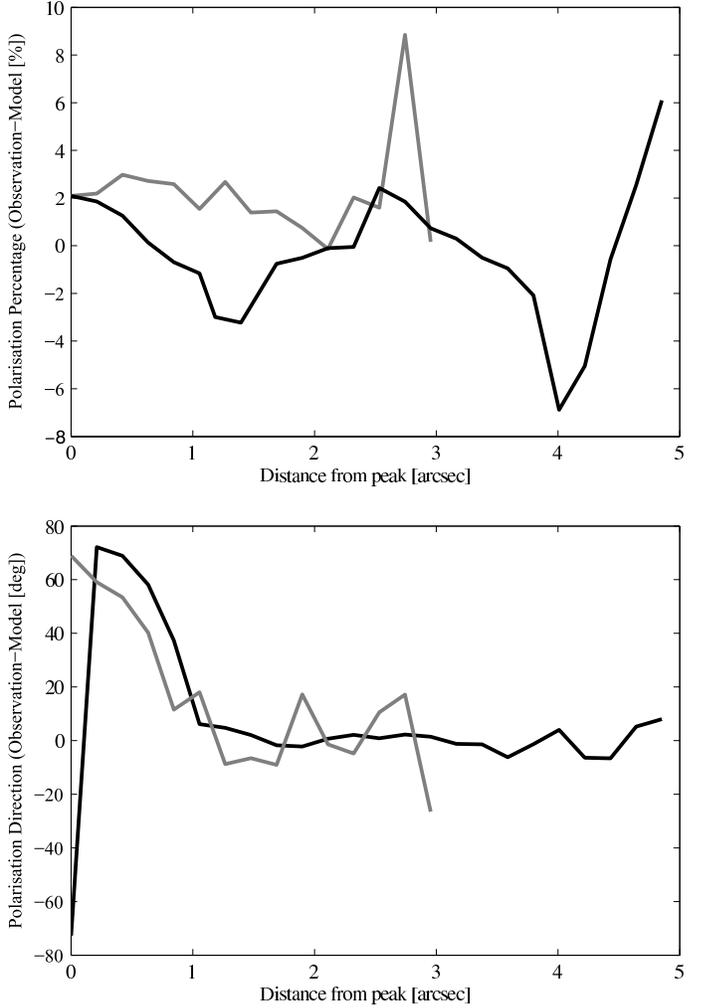}}
\caption{Difference of observed and modelled polarisation percentage
 ({\sl top}) and polarisation direction ({\sl bottom}) along the
ridges (black) and the symmetry line (grey).} \label{poldiff}
\end{figure}

For the best-fitting model we compared the polarisation map with the
data. Observational effects are applied to the model polarisation
output in a similar manner as for the direct images.

To compare the model with the data, the polarisation values are
extracted along the ridges and the symmetry axis of the Northern
nebula (see Fig.~\ref{polarisation} left and middle panel) and
compared as a function of distance from the central source (right
panel). The difference of the observed and the modelled polarisation
rate and orientation can be found in Fig.~\ref{poldiff}. The
observed polarisation rates are reproduced reasonably well along the
ridges by the model ($\Delta P/P\approx 0.3$) except on the central
peak where the model predicts only a 1.2\% polarisation instead of
the observed 3.3\%. The large difference in polarisation orientation
close to the central peak is an artifact from the exact peak
placement and has no physical meaning. The upper limit of the
observed polarisation is also well reproduced ($\sim$30\%). Along
the symmetry axis of the nebula the general trend is also correct,
rising from central source to edge, but the model polarisation
appears to be systematically lower by a few percent compared to the
data. Nonetheless, the trends are well reproduced as well as the
observed maximum polarisation rates.

\begin{table}[h]
\caption{Extracted observables of the observation and the best-fit
model} \label{Par_extract} \centering
\begin{tabular}{l c c c}
\hline\hline
Observable & Band & Observation& Model\\
\hline
                    &I&15.8&14.9\\
w [arcsec]          &H&14.1&14.0\\
                    &K&10.8&12.1\\
\hline
                    &I&129&136\\
$\delta$ [$\degr$]  &H&129&133\\
                    &K&136&139\\
\hline
                    &I&0.31&0.33\\
$t$                 &H&0.44&0.33\\
                    &K&0.53&0.23\\
\hline
                    &I&340&300\\
Contrast            &H&$>$1100&-\\
                    &K&$>$8700&-\\
\hline
c [arcsec]          &I&4.76&4.22\\
\hline
&0.7$\mu m$&0.18$^1$&0.25\\
&0.9$\mu m$&0.90$^1$&1.11\\
&1.25$\mu m$&1.96$^2$-2.09$^3$&2.26\\
&1.6$\mu m$&2.24$^4$,2.9$^6$,3.42$^3$&2.85\\
Flux&2.22$\mu m$&3.06$^4$,3.46$^6$,4.31$^3$&3.61\\
$[10^{-14}Wm^{-2}$Hz$^{-1}]$&3.8$\mu m$&1.19$^2$&4.75\\
&12$\mu m$&6.50$^3$&7.19\\
&25$\mu m$&8.63$^3$&6.93\\
&60$\mu m$&7.39$^3$&6.63\\
&1300$\mu m$&0.025$^5$&0.008\\
\hline
min. Pol.           &I&3.3\%&1.2\%\\
max. Pol.           &I&30.9\%&30.3\%\\
\hline
\end{tabular}
\flushleft References: 1 -- Strom~\&~Strom (1994); 2 --
Luhman~\&~Rieke (1998); 3 -- Kenyon et al. (1990); 4 -- Luhman
(2000); 5 -- Motte~\&~Andr\'e (2001); 6 -- This paper, see
Sect.~\ref{aperture_photometry}\\
\end{table}

\section{Discussion}\label{discussion}

\subsection{Reliability of the models}

We were able to fit a dust disc model to the observations of
IRAS~04158+2805 that matches the I-, H- and K-band images as well as
the SED from visual to far-infrared and the I-band polarisation rate
at different object positions. Since our model uses simple
assumptions, such as homogeneous spherical grains, power-law
distributions (grain size, surface density, flaring) and no dust
settling, these results provide a useful insight on the geometry of
the circumstellar environment.

We fitted the observation well in terms of the width of the nebula,
opening angle and peak-to-peak distance and brightness contrast (see
Table~\ref{Par_extract} for all observables). The triangularity $t$
is well fitted for the I- and H-band but for the K-band the model is
too triangular. At the centre of the object, the model predicts a
polarisation rate which is slightly below the observed value. While
this probably indicates that our grain model needs refinement
(either in its size distribution, composition or grain shape), the
former may indicate that we need a more complex geometry either for
the disc or for the emitting source. For instance, K-band emission
from the inner parts of the disc might be an important source of
photons, which is neglected in our calculations. To check this
possibility we calculated a K-band model image that includes the
disc emission as a source for scattered photons for our best-fitting
model. It turns out that the star is so cool that, at the 0.5~AU
assumed inner radius, the dust reaches a temperature of only 400~K.
As a consequence, roughly 99\% of all K-band photons emitted by the
system come from the star and, therefore, the resulting images are
unchanged if we include or neglect the disc emission. Vertical
settling could also play a role, as well as the presence of a
remnant halo that could be responsible for the roundish 4-5"
structure best highlighted at K-band, where the extinction of the
material located in front becomes much lower.

By comparing the I-band observed and modelled images, the shape of
the counter nebula and the dark lane between the two bright areas do
not coincide exactly. The model did not take into account the
absorption of {\sl background} light in the dust lane. The
artificial noise of the model images was produced by adding Poisson
noise to the scattered light images and by superposing
position-independent Gaussian-noise. This does not reflect the real
nature of the background since IRAS~04158+2805 is illuminated by the
large reflection nebulosity in the back. Therefore, we may expect a
difference of the observation and the models at those wavelengths
where the background light dominates the noise. In the H- and
K-bands this dominance is not visible in the observed images.

The model underpredicts the millimeter flux by a factor of 3 but
uncertainties are large regarding dust opacities in this regime,
typically by a factor of 5. Also, the mm-data were obtained with an
11 arcsec beam and may suffer from background contamination, hence
overestimate the true flux. Higher resolution millimeter data is
needed to investigate this discrepancy further.

Nevertheless, one single model can describe accurately the scattered
light images in three wavelength bands, the SED and the polarisation
map of IRAS~04158+2805. The method applied here is therefore
generally valid and promising to find parameters for relatively
simply structured dust discs.

\subsection{Gas to dust ratio}

From our model, we can easily derive the column density
$\sigma_{dust}$ by integrating the dust density structure along our
line of sight to the central source. We find
$\sigma_{dust}=3.3^{+1.8}_{-1.2}\times10^{-4}$ g cm$^{-2}$. From
$N_H$ we obtain a gas column density $\sigma_{gas}=
7.2^{+4.8}_{-3.0}\times10^{-2}$ g cm$^{-2}$ by assuming interstellar
abundances (Morrison \& McCammon 1983). This provides an estimate of
the total gas-to-dust ratio in a protoplanetary disc, along the line
of sight that grazes the disc top layers in this case. To our
knowledge this is the first time it is obtained directly using $N_H$
rather than $N_{CO}$:
\begin{equation}
\frac{\sigma_{gas}}{\sigma_{dust}}=220^{+170}_{-150}
\end{equation}

While uncertainties on this ratio are still substantial, this
illustrates a robust method to derive dust-to-gas ratios in
protoplanetary discs. Although the value we find is compatible with
the standard value of 100 usually assumed in this context, our
result might suggest a slightly higher value in the top layers of
the disc. This is a potentially important result that calls for
further investigations in more sources.

In any case, we do find an upper limit to the gas-to-dust ratio
(albeit not a lower limit) of $\approx 560 (2\sigma)$, which
provides clear evidence that the disc cannot be strongly (e.g. by an
order of magnitude) depleted of dust along the line of sight to the
star. Such dust depletion would be expected if strong settling of
dust toward the disc midplane had occurred.

\subsection{IRAS~04158+2805: a classical T~Tauri star with a large disc?}

We have shown that we can well reproduce all observational
properties of IRAS04158 with a simple model of a nearly edge-on
disc, i.e., that of a Class II source without substantial
circumstellar envelope. This result confirms prior classifications
by e.g. Park \& Kenyon (2002) or Kenyon \& Hartmann (1995). The flat
SED of IRAS~04158+2805 can be interpreted in terms of a high
inclination to the line of sight instead of invoking a highly
embedded source, as a protostar would be. This work demonstrates the
importance of understanding the circumstellar geometry to assess the
nature of the central source. Here, an edge-on disc blocks our
direct view towards the central object, flattening the shape of the
SED. It is known that a classical T~Tauri star can have a rising
near- to mid-infrared SED. But only a few mimic an embedded source,
as is the case for IRAS~04158+2805. Most sources such as HH30 (see,
e.g., Wood et al. 2006), HK Tau B, or HV Tau C, do have a declining
near- to mid-IR SED, and the second peak from thermal emission is
found only at longer wavelengths.

Interestingly, IRAS~04158+2804 is just above the substellar limit,
based on its spectral type. If our model is correct, it hosts a
large massive disc implying that some of the lowest-mass T~Tauri
stars, at least in Taurus, can be surrounded by $\sim$1000AU-radius
discs. This object likely formed from the collapse of a small
pre-stellar core and is very unlikely to have undergone any violent
dynamical interaction, such as an ejection from an unstable multiple
system (e.g. Reipurth \& Clarke 2001). This result adds support to
the idea that the ejection scenario is not the only mode to form VLM
T~Tauri stars and brown dwarfs.

Also, considering the probable mass of the central star $\sim
0.1-0.2 M_\odot$, the total dust-disc mass we derive
($1-2\cdot10^{-4} M_\odot$) and the gas-to-dust ratio we derived
($\sim$220, assuming it is true throughout the disc and not only
along our line of sight), this implies a total disc/star mass ratio
of $\sim$0.1--0.2. In other words, the disc is close to the limit
against gravitational instability. Because collapse in these
gravitationally unstable discs is one suggested mode for planet
formation, it is of great importance to study IRAS~04158+2805
further: The estimation of both the gas and the dust masses need to
be refined. The gas mass estimation can be improved by much deeper
X-ray observations while the dust mass can be obtained better with
models witch use more free parameters and which are directly fitted
on the images, not just a handful of observables.

\section{Conclusion}\label{conclusion}

In this paper we presented a multiwavelength study of
IRAS~04158+2805 and its circumstellar environment. We modelled the
shape and brightness profiles of the reflection nebulosity in three
optical and NIR bands (I-, H- and K-band) with MCFOST, a Monte Carlo
polarised radiative transfer code, and found a good agreement
between the model and the data. Many parameters of the final model
are well constrained (e.g., inclination, $a_{\mathrm{max}}$,
$M_{\mathrm{dust\,disc}}$) while a few remain poorly determined. The
scale height h and $\beta$ are degenerate but the pair h-$\beta$ is
relatively constrained.

The disc model parameters used to match the data are listed in
Table~\ref{Bestfit}. They are 2$\mu m$ for the maximum dust grain
size, 1.5$\cdot10^{-4}M_{\odot}$ for the dust disc mass, 1120 AU for
the outer disc radius, 1.2 for the exponent $\beta$ of the flaring
law, -1.0 for the exponent $\alpha$ of the radial dust density law,
a scale height of 8AU at a radius of 50 AU and an inclination of
62.7\degr.

The best model fits the observed SED well. However, it falls
slightly short of producing the right amount of 1.3mm continuum flux
by about a factor of 3. The model also reproduces reasonably the
observed I-band polarisation behaviour along the symmetry axis and
ridges of the Northern nebula.

Combining dust disc models with X-ray spectroscopy allowed probing
the gas-to-dust ratio in the disc. We found a value of
$220^{+170}_{-150}$ which is compatible with the ISM value and the
value generally assumed in protoplanetary discs.

According to its spectral type, IRAS~04158+2805 has a mass slightly
above the substellar limit. Clearly, stars with such a low mass keep
being formed by accretion from a circumstellar disc whose properties
do not seem to differ significantly from those of their more massive
young counterparts. It would be interesting to push the search
further for disc images around less massive objects, located well
into the substellar regime.

\begin{acknowledgements} This work is based in part on archival
data obtained with the {\sc Spitzer} Space Telescope, which is
operated by the Jet Propulsion Laboratory, California Institute of
Technology under a contract with NASA. Support for this work was
provided in part by an award issued by JPL/Caltech and in part by
{\sl Programme National de Physique Stellaire} (PNPS) of CNRS/INSU
(France). We are grateful to Sylvain Bontemps for obtaining the
CFHTIR images presented here and to Jerome Bouvier for performing
their basic data reduction. The CXC X-ray Observatory Center is
operated by the Smithsonian Astrophysical Observatory for and on
behalf of the NASA under contract NAS8-03060. \end{acknowledgements}

\end{document}